\documentclass[journal]{IEEEtran}
\ifCLASSINFOpdf
   \usepackage[pdftex]{graphicx}
\else
   \usepackage[dvips]{graphicx}
\fi
\usepackage{amsmath}
\begin{document}
%
\title{A Closer Look at Weak Label Learning \\ for Audio Events}
%
%
%

\author{Ankit Shah$^*$, Anurag Kumar$^*$\thanks{$^*$First two authors contributed equally.}, Alexander G. Hauptmann~\IEEEmembership{}
        and Bhiksha Raj ~\IEEEmembership{Fellow, ~IEEE}}
\maketitle

\begin{abstract}
Audio content analysis in terms of sound events is an important research problem for a variety of applications. Recently, the development of
weak labeling approaches for audio or sound event detection (AED) and availability of large scale weakly labeled dataset have finally opened
up the possibility of large scale AED. However, a deeper understanding of how weak labels affect the learning for sound events is still missing from literature. In this work, we first describe a CNN based approach for weakly supervised training of audio events. The approach follows some basic design principle desirable in a learning method relying on weakly labeled audio. We then describe important characteristics, which naturally arise in weakly supervised learning of sound events. We show how these aspects of weak labels affect the generalization of models. More specifically, we study how characteristics such as \emph{label density} and \emph{corruption of labels} affects weakly supervised training for audio events. We also study the feasibility of directly obtaining weak labeled data from the web without any manual label and compare it with a dataset which has been manually labeled. The analysis and understanding of these factors should be taken into picture in the development of future weak label learning methods. \emph{Audioset}, a large scale weakly labeled dataset for sound events is used in our experiments.
\end{abstract}

\begin{IEEEkeywords}
Audio Events, Weak Labels, Weakly Supervised, CNN
\end{IEEEkeywords}

%
\IEEEpeerreviewmaketitle

\section{Introduction}
%
%
%
%

\IEEEPARstart{M}{ultimedia} data on the internet is growing at an exponential rate and according to one estimate \footnote{https://www.cisco.com}, around three-quarter of internet traffic currently consists of videos. It is expected to grow more in coming years. This clearly demands methods which can automatically analyze the contents of these multimedia data. Automatically analyzing multimedia data is necessary for their successful indexing, search and retrieval as required by the end application. Audio Event Detection (AED) or Sound Event Detection deals with the problem of automatically analyzing the \emph{audio content} of these multimedia data with respect to different sound events. Besides being important for content based search and retrieval of videos on the web, AED plays a vital role in other areas as well. Sounds are crucial to our understanding of the environment around us, and the next generation of artificial intelligence systems requires capabilities to recognize, understand and interpret sounds just as humans can. Hence, AED is essential for improving human-computer interaction as well. Voice-based smart assistants such as Google Home, Amazon Alexa, Apple Siri, etc. can draw context-related information using AED\footnote{http://blog-idcuk.com/sound-recognition-as-a-key-strategic-technology-for-artificial-intelligence/} and can improve their interaction with users by using such information. Another important application is surveillance. The importance of audio based surveillance has been known and studied for a long time \cite{clavel2005events}. Audio does not require ``line of sight" and are relatively easier to capture and transmit for surveillance purposes.   

The development of large-scale audio event detection systems has been hindered for a long time by the availability of a large labeled dataset. The problem primarily lies in annotating audio recordings for sound events. Annotating audio recordings with time stamps of occurrences of sound events is a difficult, time consuming and an expensive task. This is evident in the small scale of several sound event datasets and challenges released in the last decade or so \cite{stowell2015detection, mesaros2016tut, zieger2005acoustic, piczak2015esc, salamon2014dataset}. The scale here refers to both (1) amount of available training data for each event and (2) vocabulary of sound events. Labeled audio recordings with time stamps are often referred to as \emph{strongly labeled} data \cite{Kumar:2016:AED:2964284.2964310}. Hence, supervised learning methods which rely on event exemplars for training, or in other words time stamped information to extract out event exemplars, is not expected to be a scalable approach to AED, as large amounts of strongly labeled data is very hard to come by. 

Audio event detection using weakly labeled data, proposed in \cite{Kumar:2016:AED:2964284.2964310}, addresses the problems of strongly labeled data. In \emph{weakly labeled} audio data, only the presence of absence of an event is known. That is an audio recording is just tagged with respect to an event. For example, for an audio recording which might be several minutes long, we just know whether a sound event, say \emph{dog barking} occurred or not. No other information such as time stamps or count of occurrences of \emph{dog barking} within the recording is known.  Hence for weak labels, manually tagging audio recordings with weak labels is much easier compared to annotating them with timestamps of events. Moreover, as highlighted in \cite{kumar2017deep}, weak labels also address ambiguities or interpretation issues (faced by annotators), in defining the boundaries of sound events. It also opens up the possibility of training AED systems from the vast amount of audio (or video) data available on the web. Metadata associated with web audios (or videos) can be used to obtain weak labels for audio recordings, which can further reduce labeling effort. 

Overall, learning audio events using weakly labeled data is a promising paradigm to scale audio event detection. Several methods have been proposed for weakly supervised training of sound events. AED with weak labels is now a task in the annual IEEE AASP sound event and scene classification challenge as well (DCASE 2017\cite{mesaros2017dcase}, DCASE 2018 \footnote{http://dcase.community/challenge2018/}). \emph{Audioset} \cite{AudioSet45857}, a large scale weakly labeled audio event dataset have further helped accelerate research in this direction. 

However, learning sound events from weakly labeled data comes with its own set of problems and challenges. Although, several methods (see related work section) have been proposed for AED using weak labels in the last couple of years, a deeper understanding and analysis of the problems and challenges arising in large-scale AED using weak labels is entirely missing. In this paper, we take a closer look at AED using weakly labeled data. We first describe a Convolutional Neural Network (CNN) based approach for large-scale audio event detection using weak labels. We then propose different ways in which ``label noise" naturally manifests itself into weak label paradigm for sound events. These characteristics of weak label learning are then analyzed to understand how they affect the generalization capabilities of the trained model. This analysis and understanding is not only desirable but necessary for further development of audio event detection using weakly labeled data. One needs to factor in these characteristics when developing a learning algorithm.   

The primary motivation behind weak label learning is that it shows a way to scale the learning process by easing the data acquisition and labeling process. Primarily its significance lies in the fact that it is a learning paradigm where data from the web can be exploited. However, it comes at the cost of noise in the training data. This aspect of weakly supervised learning has been explored in the field of computer vision, \cite{lu2017learning}\cite{tang2009inferring}\cite{feng2014image} to cite a few. 

However for sounds, weakly supervised learning has only recently come up and the impact and relevance of factors such as noise in the data is yet to be analyzed, understood and explored. This paper is an attempt in that direction. For sounds, weakly labeled data from the web (such as YouTube) contains noise primarily in two forms, \emph{Signal noise} and \emph{Label noise}. We use the term \emph{signal noise} to refer to a variety of signal degrading factors. Often, consumer-generated web audios or videos are recorded under unstructured conditions. Occurrences of sound events of interests are often corrupted by either noise and/or events of which are not of interest. The recording conditions, environments, styles, instruments can all lead to huge intra-class variations. This does make the problem much more challenging. The focus of this work, however, is on analyzing the other type noise in weakly supervised learning of sounds, namely, the \emph{Label noise}. 

\emph{Label noise} in weakly labeled audio occurs naturally in different ways. In this work, we consider two important factors. The first one is the \emph{Label Density}. Label density signifies the portion  of the recording (out of the total length) during which the tagged sound event is actually present in the recording. Consider an example. We have an audio recording ($R_1$) of 10 seconds containing only a total of 1 second of the tagged event in it and another audio recording ($R_2$) of the same duration in which the tagged event is present for a total of 5 seconds. Clearly, the density of labeled event in $R_2$ is higher. In other words, $R_1$ in its entirety, is a weaker representation of the event compared to $R_2$. We can define label density noise as a measure of \emph{weakness} of a given audio recording with respect to the tagged audio event.

The second form of \emph{label noise} comes from the actual wrong labeling of the audio recordings. Sound events can often be hard to interpret, and this might lead to wrong labels, even when the labeling is done manually. This becomes a bigger problem when we work on large scale, with a large number of audio events. For example, \emph{Audioset} which have been manually labeled; the quality of the labels has been roughly estimated to be less than $90\%$\footnote{https://research.google.com/audioset/dataset/index.html} for several events. The problem can grow by several folds when we try to mine the data directly from the web and obtain weak labels by exploiting the associated metadata. In this case, a considerable portion of the weak labels is expected to be wrong. We refer to this form of label noise as \emph{label corruption noise}. 

Clearly, \emph{label density noise} is an inherent characteristic of weak label learning. Corrupt weak labels can also easily occur and hence often the two forms of noises are simultaneously present. This is especially true for weakly labeled data obtained from the web in which no manual labeling effort was employed. We try to understand these factors by empirically analyzing the performance of a state-of-art CNN based method for weakly labeled audio event classification. We first describe our CNN based approach for weakly supervised training of sound events in Section \ref{sec:wcnn}. We then describe the different label noises in details and the corresponding experimental design to analyze them in Section \ref{sec:expdesign}. We discuss our experiments and results in Section \ref{sec:experiments} and then conclude in Section \ref{sec:conclusions}. An overview of related works is given in the next section. 

\section{Related Work}
Some of the earliest works on automated content analysis of audio recordings were done for content based retrieval of audio recordings \cite{wold1996content}\cite{guo2003content}. Other earlier works focused on the detection of specific audio events such as gunshots and screams for surveillance purposes \cite{valenzise2007scream}\cite{gerosa2007scream}. Analyzing different type of acoustic features and time-frequency representations with different types of classifiers is central to these works. More recently,  Audio event detection research has started to receive much more attention, primarily due to its application in a wide range of systems; beyond the obvious content-based retrieval of multimedia data and surveillance. Human-computer interaction \cite{maxime2014sound}, smart devices including voice-based assistants \footnote{http://blog-idcuk.com/sound-recognition-as-a-key-strategic-technology-for-artificial-intelligence/}, smart homes and health monitoring systems \cite{debes2016monitoring}\cite{greene2016iot}\cite{zigel2009method, li2010acoustic}, wildlife monitoring \cite{stowell2014automatic}, all require automated audio content analysis in terms of sound events. 

Several fully supervised learning methods have been proposed in the last few years for audio event detection. For example, \cite{zhuang2010real} used a GMM-HMM architecture similar to speech recognition for AED. Bag of audio words representation built over different features is a popular and widely used representation, often combined with different classifiers\cite{pancoast2012bag,lim2015robust, lu2014sparse, kumar2012audio}. Taking cues from sound source separation where Non-Negative Matrix Factorization (NMF) have been very successful, NMF has been exploited in a variety of ways for audio event detection \cite{gemmeke2013exemplar, heittola2011sound}. More recently several deep learning methods, especially using Convolutional Neural Networks \cite{piczak2015environmental, zhang2015robust, phan2016robust, hershey2017cnn} have been proposed for audio event classification and detection. However, as pointed out earlier, fully supervised learning of sound events cannot be scaled due to difficulties in creating large-scale \emph{strongly labeled} datasets.  

\cite{Kumar:2016:AED:2964284.2964310} first proposed weak label learning for generic sound events. The authors formulated it as a Multiple Instance Learning (MIL) \cite{zhou2004multi} problem and used Support Vector Machine (SVM) and feedforward neural network based MIL methods as learning algorithms. Note that, although the training is done using weakly labeled data where temporal information is not available, the authors showed that temporal localization is possible.  MIL methods often suffer from scalability issues, especially iterative methods such as miSVM \cite{andrews2003support} in which an SVM is solved at each iteration. \cite{kumar2016weakly} proposed scalable MIL methods for audio event classification using weak labels. The primary idea is to embed \emph{bags} into vector representations and change the hypothesis space to full supervised methods. In this process, however, temporal localization capabilities are lost. 

Several deep neural networks based methods have been recently proposed for weakly supervised audio event detection. A significant chunk of these methods use convolutional neural networks \cite{su2017weakly, kumar2017deep, xu2017attention, kumar2017knowledge}. \cite{kumar2017deep} is based on the idea that one can design a CNN to scan and produce outputs at small segments (say 1 second) and then map these segments level outputs to full recording level outputs. Once this mapping is achieved one can compute loss and train the network, as weak labeling gives labels at full recording level. This idea also forms the basis of the approach taken in this work. \cite{su2017weakly} operates at frame level and combines CNN with event-specific Gaussian filters to produce recording level outputs before computing loss. \cite{xu2017attention} combines CNN with recurrent architecture to incorporate attention in the learning framework. CNN's operate on small (32 milliseconds) duration windows and the outputs from these CNN's are fed into a recurrent neural network to produce recording level outputs for computing loss. Although the idea of having a recurrent architecture following a CNN seems promising, the proposed method in the current form is not feasible for long duration (say several minutes long) audio recordings efficiently. Our CNN based method used in this work is capable of handling long and varying duration audio recordings. Another work \cite{hershey2017cnn}, ignores the weakly labeled nature of web data and trains well known CNN architectures such as VGG \cite{simonyan2014very}, ResNet \cite{he2016deep} by making strong label assumptions. However, \cite{kumar2017knowledge} shows that it is beneficial to treat weakly labeled data as such. 

Weak label learning for sound events has also been now incorporated as a task in the annual IEEE challenge on sound events and acoustic scenes (DCASE 2017 \cite{mesaros2017dcase}, DCASE 2018 \footnote{http://dcase.community/}). A small subset of Audioset is used in the challenge. Most of the top performing methods on the weak label task in DCASE 2017 \footnote{http://www.cs.tut.fi/sgn/arg/dcase2017/challenge/task-large-scale-sound-event-detection-results} relied on convolutional neural networks. 

The central theme of \emph{noise} in weak label learning paradigm for sounds is completely missing from the literature to the best of our knowledge. Perhaps the closest related work in this respect is \cite{kumar2017audio}. \cite{kumar2017audio} acknowledges the theme of \emph{noise} in weakly supervised learning and proposed that it can be addressed by a unified learning framework which learns simultaneously from strongly and weakly labeled data. The authors proposed that this unified learning framework can be formulated as a constraint form of semi-supervised learning. A graph-based constraint semi-supervised learning method is then proposed as the solution. However, a more in-depth analysis; showing how \emph{label noise} in different forms can manifest itself in weakly supervised learning has not been undertaken. Moreover, the proposed graph-based is hard to scale for a large amount of audio data. For example, we consider experiments on over 360 hours of audio data. The constraint optimization problem described in \cite{kumar2017audio} work can be very difficult to solve on such large datasets.  

\section{CNN for Weakly labeled audio}
\begin{figure*}[t!]
  \includegraphics[width=1.0\textwidth]{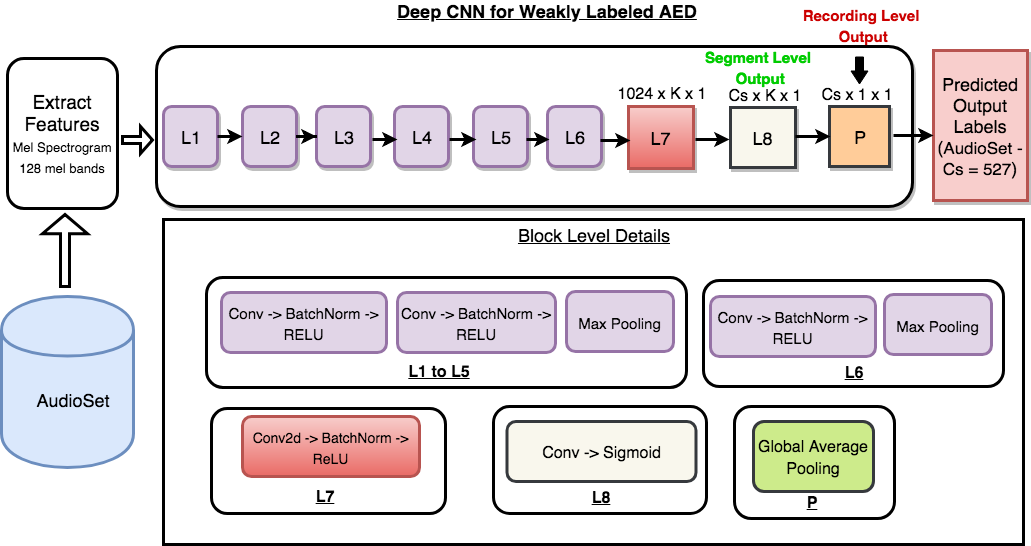}
  \caption{Weak Audio Label (WAL-Net) for Audio Events. The layer blocks from L1 to L6 consists of 2 or 1 convolutional layers followed by max pooling. L8 represents segment level output which is further mapped by the global average pooling layer P to produce recording level output. The loss is then computed with respect to the target recording level labels.}
  \label{fig:network_architecture}
  \vspace{-0.1in}
\end{figure*}
\label{sec:wcnn}
Our CNN for weakly labeled AED is designed for logmel spectrograms as input. An audio recording is represented by $X \in R^{n \,\, \times \,\, m}$ logmel spectrograms. $n$ is the number of logmel frames and $m$ is the number of Mel-filters used in the representation. For the purposed of discussion in this section, we will use $m = 128$ and hence an audio recording is represented by a time-frequency matrix $X \in R^{n X 128}$. We will use the term ``frame" to refer to the $i^{th}$ $128$ dimensional time frame, ``segment" to represent a small segment of the whole audio recording. In terms of logmel features a segment is a collection of $k$ continuous frames ($k \times 128$ matrix). For example, $k=128$ represents can represent a $1.5$ second segment. The details of logmel features are provided in experiments section. 

Given a collection of weakly labeled recordings ($X_i, y_i$), the goal is to design a network architecture which can learn from this weakly labeled dataset. We would expect the network to factor in the fact that the labels are \emph{weak}. We have a label for the whole recording but that labeled event need not be present in the whole recording. Moreover, the network should be able to handle audio recordings of variable length as the recordings can be a few seconds long or may be even a few minutes long. The primary idea here is to design the network such that we can predict posteriors of the presence of the events at small segments (say 1 or 1.5 seconds), and then map these segment level posteriors to recording level posteriors. Once we have the recording level posteriors we can compute the loss with respect to the available recording level labels and then train the network using backpropagation.    

\subsection{CNN Architecture}
The ideas outlined in previous paragraph are conceptualized by the network shown in Figure \ref{fig:network_architecture}, referred to as WAL-Net.  The blocks of layers L1 to L5 consists of two convolutional layers followed by a max pooling layer. The convolutional layers consists of batch normalization before non-linear activation function. ReLU ($max(0,x)$) \cite{nair2010rectified} is used in all layers from L1 to L6. $3 \times 3$  convolutional filters are used in all layers from L1 to L6. Stride and padding values are fixed to $1$. The number of filters employed in different layers is as follows, \emph{\{L1: 16, L2:32, L3:64, L4:128, L5:256, L6:512 \}}. Note that L1 to L6 consists of two convolutional layers followed by max pooling. Both convolutional layers in these blocks employ same number of filters. The max pooling operation is applied over a window of size $2 \times 2$. Logmel inputs are treated as single channel inputs. 

The convolutional layers in L1 does not change the height and width of input. The max pooling operation reduces the size by a factor of $2$. For example, a recording consisting of $128$ logmel frames that is an input of size $X \in R^{1 \times 128 \times 128}$, will produce and output of size $16 \times 64 \times 64$ after L1.  This design aspect applies  to all layer blocks from L1 to L6 and hence $1 \times 128 \times 128$ input will produce a $512 \times 2 \times 2$ output after L6. L7 is a convolutional layer with $1024$ filters of size $2 \times 2$. Once again ReLU activation is used. Stride is again fixed to $1$ and no padding is used. 

Layer L8 represents segment level output. It is a convolutional layer consisting of $C_s$ filters of size $1 \times 1$. $C_s$ is the number of class in the dataset. Sigmoid non-linear activation function is used at this secondary output layer. The final \emph{global pooling player}, $P$, maps the segment level outputs from L8 to full recording level outputs. In this case we use \emph{average} function to perform this mapping. That is the recording-level posterior for each class is the average of the segment-level posteriors for each class. 

\subsection{Characteristics of WAL-Net}
This network design embodies the different aspects of weakly labeled audio data outlined at the beginning of this section. The network is fully convolutional which allows it to handle recording of variable length. It is designed to predict posteriors at recording level by first predicting class posteriors at small segments. The principle applied here is that we scan through the recording and predict outputs at every small segment. Once we know what happened at segment level, we can use it to assign what will happen at recording level. In principle, this segment to recording level mapping of posteriors can be done by a different functions. For example, one can also use \emph{max} function. This would imply that the recording level posterior for a class is the highest probability of occurrence of that event across different segments. We use \emph{average} because empirically we found that it works better than max. It is possible to design other complex mapping function as well. 

The network design controls the size of the segments over which posteriors are predicted. The amount by which the segment moves can also be controlled by the network design as per requirement. In the specific case of WAL-Net presented in Fig \ref{fig:network_architecture} (and used in this work), the segment size is 128 frames ($\sim$ 1.5 seconds) and the segment can be thought of as moving by 64 frames ($\sim$ 0.75 seconds). Hence, consecutive segments overlap by $50\%$. The number of segments obtained after layer L8 depends on the duration of the audio recording. For example, one can easily verify that an audio recording consisting of $864$ logmel frames ($X \in R^{864 \times 128}$) will produce $K=12$ segments at L8. If needed these segment level outputs can give temporal localization of the event as well. Hence, the network is capable of localizing events despite learning from weakly labeled data. 

\subsection{Multi Label Training for WAL-Net}
An audio recording can have multiple classes present in it. The weakly labeled \emph{Audioset}, on an average consists of $2.7$ labels per recording. Hence, the training procedure needs to consider presence of multiple labels in the recording. The output of WAL-Net gives class specific posteriors for any given input. The binary cross entropy loss with respect to each class is then given by Eq. \ref{eq:bceloss}. 
\begin{equation}
\label{eq:bceloss}
l(y_c,p_c) = -y_c*log(p_c) - (1-y_c)*log(1-p_c)
\end{equation}
 $y_c$ is the target output for $c^{th}$ class, $1$ if $c^{th}$ event is marked to be present and $0$ otherwise. $pc$ = WAL-Net$(X)$ is the network output for the $c^{th}$ class. The training loss is the mean of losses over all classes 
 \begin{equation}
 L(X,y) = \frac{1}{C} * \Sigma_{c=1}^{C_s} \,\,\, l(y_c,p_c). 
 \end{equation}

\section{Label Noises and Corresponding Experimental Designs}
\label{sec:expdesign}
\subsection{Labels Density}
\label{ssec:labden}
Label density signifies how much of a given audio recording actually contains the the tagged event. In weakly labeled data, we only know whether an event is present or not. We do not know what portion of the recording actually contains the marked event. Hence, the concept of \emph{label density} and correspondingly  \emph{label density noise} is naturally embodied into weakly label learning. We formally define label density (LD) of a recording $R$, with respect to an audio event $e$ as
\begin{align}
LD_R^e & = \frac{Duration \, (seconds)\,\,\, for\,\, which\,\, e\,\, is present\,\, in \,\, R }{Length\,\, (seconds) \,\, of \,\, R}
\end{align}
Correspondingly, label density noise (LDN) is defined as $LDN_R^e = 1 - LD_R^e$. Label density noise is a measure of \emph{weakness} of a recording with respect to a given event. In other words, how weak are the labels for a given weakly labeled audio recording. A one-minute long audio recording where a sound event is present for only a couple of seconds is a much more weaker representation of the event, compared to a similar audio recording where the event is present for almost three-quarters of the duration of the recording. 

Understanding how label density affects the learning process in weakly supervised paradigm is important. However, designing an evaluation strategy to measure the impact of label density on generalization capabilities of trained models is not straight forward. Any such method would require one to measure label density of in all recordings with respect to each event. This would clearly require manually obtaining the duration for which a given event is present in the audio recording. Clearly, this cannot be done for a large scale dataset such as \emph{Audioset}. Hence, we consider an alternative view on label density which allows us to empirically study its impact on weakly labeled AED. 

The alternate view is based on the intuition that if we mine weakly labeled audio from the web, then the expected density of labels for a given audio event will be lower for long duration audio recordings. In other words, on an average long duration audio recordings will have higher ``label density noise" compared to shorter audio recordings. 

Taking this alternate view, we design our experiment as follows. We consider \emph{Audioset}, a large scale weakly labeled dataset for sound events. \emph{Audioset} provides weak labels for YouTube audio recordings. Weak labeling was manually done on audio recordings of 10 seconds duration (a small fraction are of less than 10 seconds). These 10 seconds audio segments were actually obtained from longer YouTube videos. 

Let us represent the Audioset training data by tuples of form ($R^i, YT_{id}^{i}, S^{i}, E^{i}, L^{i} $). $R^i$ is the $i^{th}$ recording in Audioset and $YT_{id}^{i}$ is the YouTube id of $R^i$. $S^{i}$ is the start time of $R^i$ in $YT_{id}^{i}$ and $E^{i}$ is the end time of $R^i$ in $YT_{id}^{i}$. That is $R^i$ is a portion of the video $YT_{id}^{i}$ on YouTube, starting at $S^{i}$ and ending at $E^{i}$. $R^i$'s are weakly labeled (manually). $L^{i}$ represents the set of audio events present in $R^i$. Hence, this labeling tells us that the set of events in $L^{i}$ are somewhere present in $YT_{id}^{i}$ between $S^{i}$ and $E^{i}$. We request the readers to take a quick look at the label sets provided by Audioset\footnote{https://research.google.com/audioset/download.html} to understand this aspect. 

$R_i$'s are mostly 10 seconds long recordings and will have label densities likely on the higher side. For each $YT_{id}^{i}$ in training set, we consider the audio from $S^{i} - 10$ seconds to $E^{i} + 10$ seconds. Clearly, the set of $L^{i}$ events are present in this audio as well and hence we obtain 30 seconds long weakly labeled audio recordings. Those  $YT_{id}^{i}$ where the total duration of YouTube video is less than 30 seconds, we simply consider the whole recording. We call the obtained weakly labeled set as \emph{Audioset-At-30}. It is expected that on an average the label density of events in  Audioset-At-30 will be lower than Audioset. Similarly, we obtain \emph{Audioset-At-60} by considering $S^{i} - 25$ seconds to $E^{i} + 25$ seconds for each ($YT_{id}^{i}, S^{i}, E^{i}, L^{i} $). Once again $YT_{id}^{i}$ which are of less than 60 seconds are taken entirely.  \emph{Audioset-At-60} is expected to have even lower label density or higher label density noise. We train our proposed WAL-Net on all three sets and analyze how label density affects the performance. Since our network is designed to handle audio recordings of variable length, varying length of recordings is not a problem. 

\subsection{Corrupted Labels Noise}
\label{ssec:labcor}
Corrupted labels or wrong labels is possible due to a variety of reasons. Sound events are often hard to interpret which might present difficulties in labeling recordings even when manually done. This is especially true for YouTube quality audio recordings where ``signal noise" can create further difficulties in understanding and labeling the audio events. For Audioset, the authors  themselves checked a random sample of 10 recordings for each event and provided a confidence on correctness or quality of labels \footnote{https://research.google.com/audioset/dataset/index.html}. One can observe that this confidence is less than $90\%$ for several events. 

Quality of labels can reduce by a considerable amount when we try to directly mine audio from web and automate the weak labeling by using metadata associated with audios (videos) to assign labels. Both false positive as well as false negative assignment of labels are possible. That is we may mark an event to be present in the recording when its not. Similarly, we might miss the presence of an event and mark the recording to not contain the given sound event. Since one of the major motivation behind weak labels is the fact that we can exploit large amount of audio from web by automatically labeling them and then train large scale models; it is imperative that we take a closer look at this form of label noise. 

We analyze this form of label noise by corrupting the labels in Audioset. We treat the labels from Audioset as perfect or at $0\%$ corruption. We note that this is actually not true. However, Audioset has been manually labeled and it is the best one can possibly do for a collection of over $500$ sound events. Hence, we consider Audioset as having perfect labels without any noise. We then gradually increase the amount of corrupted labels by manually corrupting the assigned labels in Audioset. More specifically, for a portion of the labeled tuples ($YT_{id}^{i}, S^{i}, E^{i}, L^{i} $), we corrupt $L^{i}$ by changing the events in $L^{i}$. The corruption process is done in a stratified way such that around $r\%$ labels for each event get corrupted. $r\%$ includes both false positive and false negative label noises. Moreover, we perform the corruption in a way such that the number of recordings marked to contain the recording remains consistent with respect to original labels. We perform analysis on different values of $r$. The set of corrupted labels for each $r$ will be released for future works.

\subsection{Weakly Labeled Audio In the Wild}
\label{ssec:ytwild}
This part tries to understand challenges of \emph{label noise} in weakly labeled audio in a situation where both of the above forms of \emph{label noises} can occur in abundance. We directly obtain audio from YouTube for a given sound event and then train our model using these data. We apply a simple but effective filtering approach by retrieving videos based on the search queries of form  ``\textless sound event name \textgreater \hspace{1pt} sound". Adding the word ``sound" leads to a considerable improvement in retrieval of relevant videos on YouTube. 

For each event we consider the top $50$ retrieved videos (under 4 minutes duration) and mark these to contain the event. If a video is retrieved for multiple events, they are accordingly multi-labeled. We call this training set \emph{YouTube-wild}. In \emph{YouTube-wild} label noises in all forms occur. Label density can vary from $0$ to $1$. Since the retrieval is not perfect, marking all top $50$ videos to contain the event introduces false positive labeling, implying we are assigning a positive label to the recording even when the event is not present. False negative labels also gets naturally introduced in the process, as a retrieved video $V$ for an event, $e_1$, might contain another event, $e_2$, as well. But unless $V$ was retrieved for $e_2$ also, we do not know this and we mark $e_2$ to be not present in $V$.  

Several of the audio events in \emph{Audioset} have vague names and represent an extremely broad meaning. The automated labeling process described in the previous paragraph leads to more or less useless results for these events. Hence, for this part we worked with a smaller subset of events; selecting those which have more definite meaning. That is those for which retrieval leads to somewhat meaningful and relevant set of audio recordings. The selection process also factors in the total number of examples available for the event in Audioset. We select events for which more examples are available in Audioset. We do this because we wanted to analyze how \emph{YouTube-wild} compares with \emph{Audioset}, which is manually labeled. It is desirable that we work with events for which higher number of examples are available in Audioset, to better understand where \emph{YouTube-wild} training stand in comparison to Audioset. YouTube-wild is collected and labeled without manual effort and contains long duration audio recordings, against Audioset, which is manually labeled and weak labels are over relatively short 10 seconds recordings. Since, the source of both is YouTube, signal noise is expected to be similar. The two forms of ``label noise", however, are going to make the most difference. More details in experiments section.  

\begin{figure}[t!]
\centering
\includegraphics[width=0.49\linewidth,height=1.0in]{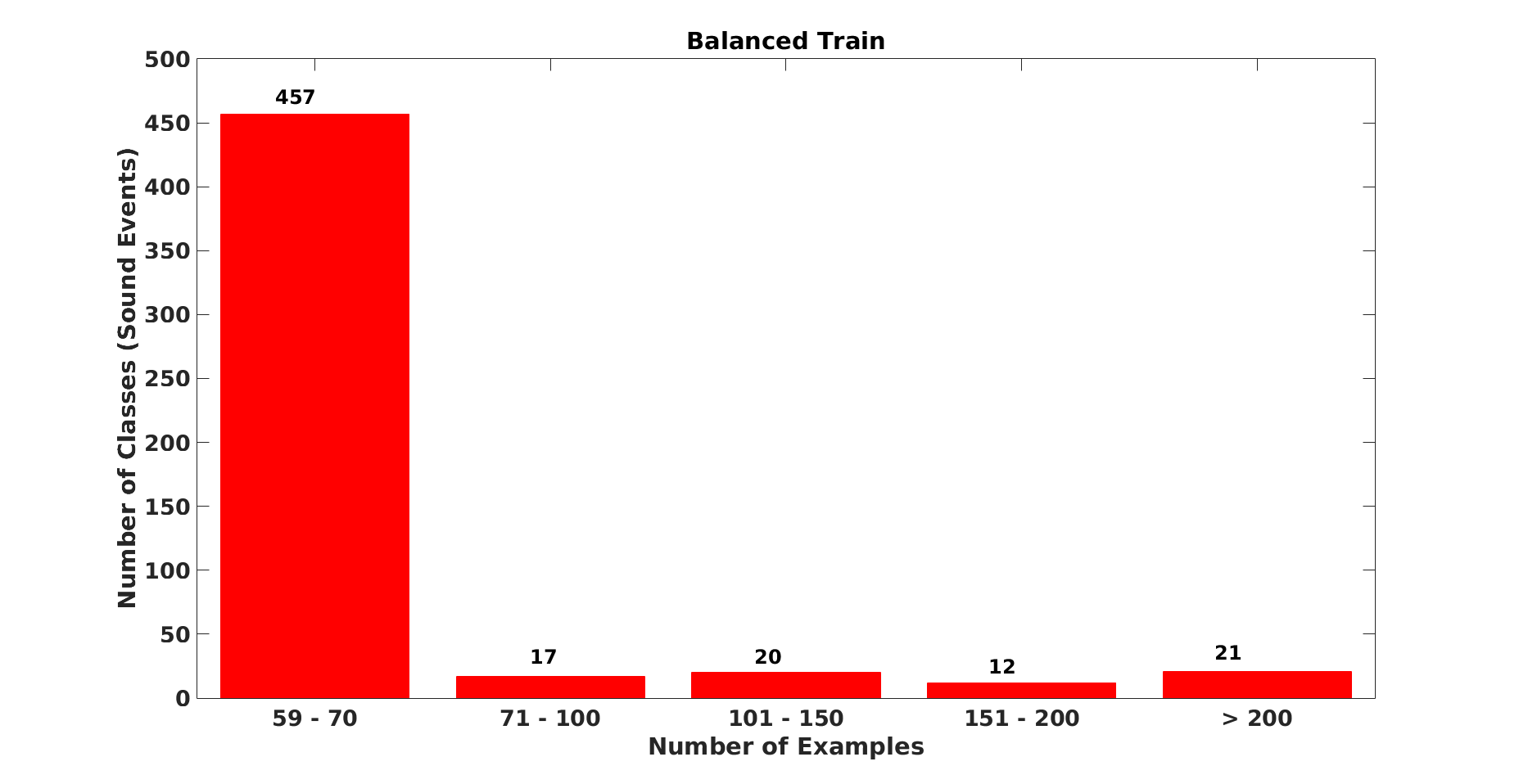}
\includegraphics[width=0.49\linewidth,height=1.0in]{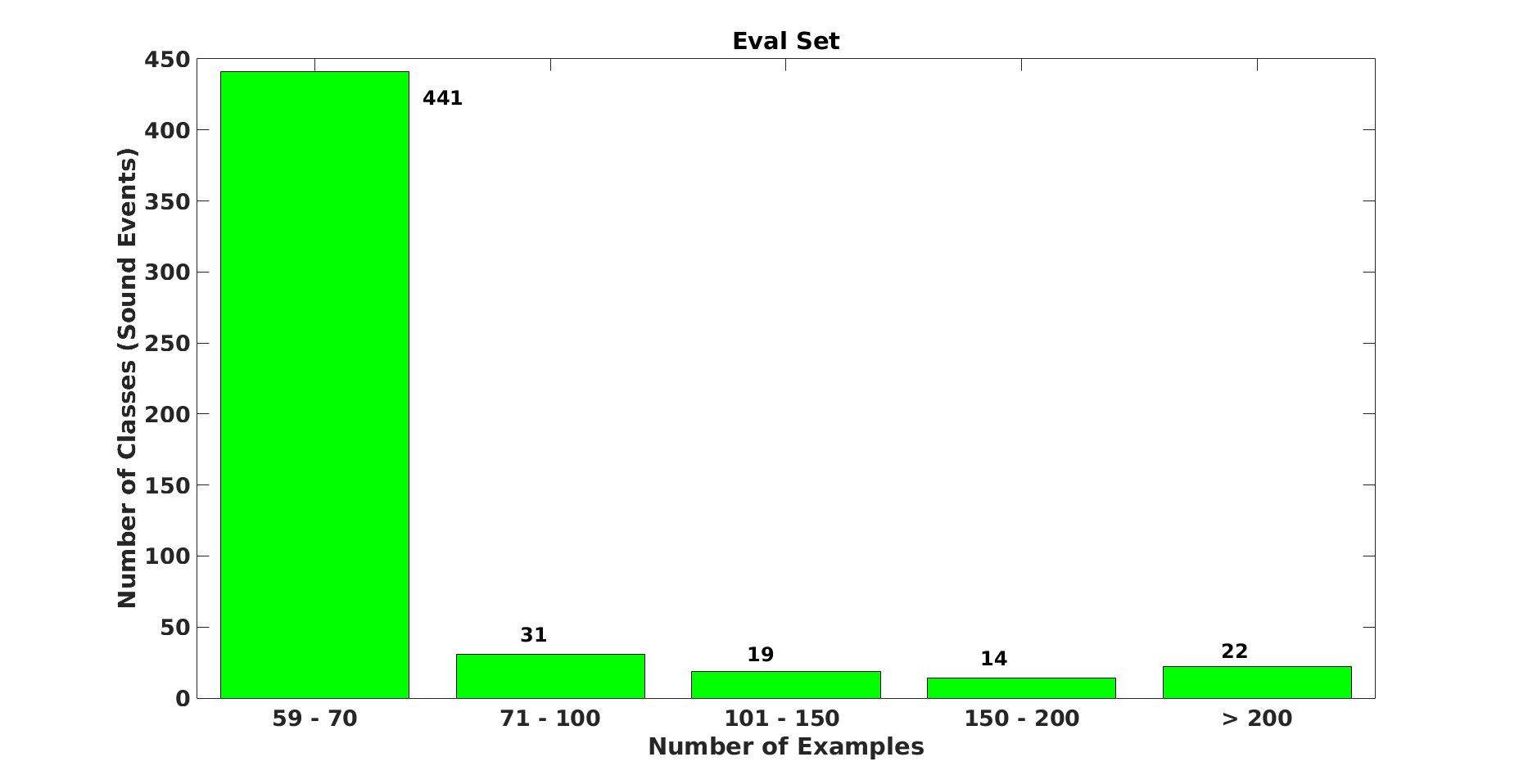}
\caption{Left: Distribution of labels in training set. Right: Distribution of labels in evaluation set. X axis shows a range of number of examples and the Y axis shows number of events for which number of examples lies in that range. For training set, out of the total 527 events, 457 events have recording counts between 59 and 70. This number is 441 for evaluation set. \vspace{-0.1in}}
\label{fig:audset}
\vspace{-0.05in}
\end{figure}

\section{Experiments and Results}
\label{sec:experiments}
\subsection{Dataset}
We use AudioSet \cite{AudioSet45857} in our experiments, a large scale weakly labeled dataset consisting of $527$ sound events. The dataset comes pre-divided into $3$ sets. We use the \emph{balanced training} set as the training set for our experiments. The balanced training set consists of a total of 
21508 audio recordings. The actual balanced training set available from Audioset is slightly larger, some of the videos were not available for download at the time we downloaded them. Most of the recordings are of 10 seconds duration, although a few are less than 10 seconds as well. The total training data is approximately $60$ hours of data. We use the \emph{Eval} set from Audioset as the test set in our experiments. The \emph{Eval} set consists of a total of 19,789 recordings. The training and test sets consists of atleast 59 examples per class. We sample out a collection of 12,676 recordings from \emph{Unbalanced} set of Audioset to use as validation set in our experiments. This validation set consists of atleast 30 examples per class. Audioset is a multi-label dataset and on an average $2.7$ labels per recording are present. The distribution of labels across different events is unbalanced. Fig \ref{fig:audset} shows this imbalance. 

\subsection{Acoustic Features, Implementation and Evaluation Metrics}
We use Logmel spectrograms as acoustic features for audio recordings. All audio recordings are sampled at 44.1 KHz sampling frequency. A window of $\sim$  23ms (1024 points) and window hop size of $\sim$ 11.5 ms is used. The number of mel bands is set to 128. Hence, each frame of $X \in R^{k \, \times \, 128}$ corresponds to the logmel representation of a 32 ms window. As mentioned in Section \ref{sec:wcnn}, WAL-Net is designed to produce outputs for a 128 frame segment moving by 64 frames. This corresponds to a segment size of $\sim$ 1.5 seconds and segment hop size of $\sim$ 0.75 seconds. 

We use Pytorch \footnote{http://pytorch.org/} deep learning toolkit in our experiments. Hyper parameters such as learning rate is tuned using the validation set. Adam optimization \cite{kingma2014adam} is used for training the networks and model selection (across different epochs) and parameter tuning is done using the validation set.  

We use average precision (AP) \cite{buckley2004retrieval} and area under ROC curves for each event (AUC) \cite{fawcett2004roc} as evaluation metrics. Mean Average Precision (MAP) and Mean Area Under ROC curves (MAUC) over all events are used as overall evaluation metrics. Temporal localization performance numbers are not reported because Audioset does not provide temporal information to compute any suitable metric. 

\subsection{Audioset Performance}
\begin{table}[t!]
  \centering
  \caption{MAP and MAUC on Audioset  }
    \begin{tabular}{|l|l|l|}
    \hline
    \textbf{Model} & \textbf{AP} & \textbf{AUC} \\
    \hline
    AlexNet(BN) \cite{2017arXiv171100229W} & NA   & 0.927  \\
    \hline
    AlexNet \cite{2017arXiv171100229W} & NA   & 0.895 \\
    \hline
    \textbf{WAL Net} & \textbf{0.196} & \textbf{0.925} \\
    \hline
    \end{tabular}%
  \label{tab:baseline}%
  \vspace{-0.05in}
  
\end{table}%

\begin{table}[t]
  \centering
  \caption{Sound Events with 10 best and worst performances (selection based on AP)  }
 \resizebox{1\columnwidth}{!}{\begin{tabular}{|l|l|l|l|l|l|}
    \hline
    \textbf{Top 10 class} & \textbf{AP} & \textbf{AUC} & \textbf{Bottom 10 class} & \textbf{AP} & \textbf{AUC} \\
    \hline
    Bagpipes & 0.804 & 0.993 & Noise & 0.015 & 0.805 \\
    \hline
    Music & 0.727 & 0.890 & Harmonic & 0.015 & 0.845 \\
    \hline
    Battle cry & 0.667 & 0.997 & Burst pop & 0.015 & 0.833 \\
    \hline
    Change ringing (campanology) & 0.658 & 0.989 & Inside, public space & 0.014 & 0.782 \\
    \hline
    Speech & 0.646 & 0.864 & Buzz & 0.014 & 0.717 \\
    \hline
    Heart sounds, heartbeat & 0.633 & 0.985 & Male speech, man speaking & 0.013 & 0.799 \\
    \hline
    Harpsichord & 0.618 & 0.981 & Bicycle & 0.012 & 0.777 \\
    \hline
    Siren & 0.601 & 0.970 & Cacophony & 0.012 & 0.796 \\
    \hline
    Civil defense siren & 0.586 & 0.987 & Scrape & 0.011 & 0.773 \\
    \hline
    Purr & 0.578 & 0.968 & Rattle & 0.010 & 0.710 \\
    \hline
    \textbf{Mean} & \textbf{0.652} & \textbf{0.962} & \textbf{Mean} & \textbf{0.013} & \textbf{0.784} \\
    \hline
    \end{tabular}%
    }
    \label{tab:baseline10}
\end{table}

Table \ref{tab:baseline} shows MAP and MAUC on evaluation set over all 527 sound events. For the purpose of comparison we also show reported numbers in \cite{2017arXiv171100229W}. Note that the training and evaluation might vary slightly in these two cases as per the availability of YouTube videos when they were downloaded.  WAL-Net is able to achieve almost similar MAUC when compared to \cite{2017arXiv171100229W}. It is not possible to show the AP and AUC numbers for all 527 events. Table \ref{tab:baseline10} shows performance numbers for 10 events for which best AP are obtained and 10 events for which worst AP are obtained. Corresponding AUC values are also reported. An important worth noting from Table \ref{tab:baseline10} is that performance is very high for events which are very specific and have salient characteristics, such as \emph{music, bagpipes, siren} whereas ambiguous events such as \emph{Harmonic, Burst pop, inside-public space} are harder to understand for WAL-Net. The ambiguities in content not only makes it harder for the network to learn these events but most likely the weakly labeled examples of these events in Audioset contains higher signal and label noise. 

\subsection{Analyzing Label Density}
The experimental setup for this part follows the procedure described in Section \ref{ssec:labden}. For the training set in Audioset, we created two additional sets, namely, \emph{Audioset-At-30} and \emph{Audioset-At-60}. Audioset-At-30 and Audioset-At-60 are approximately 180 and 360 hours of audio data respectively. These two additional training sets have higher label density noise compared to original Audioset. We train WAL-Net on original Audioset, Audioset-At-30 and Audioset-At-60. The validation and evaluation sets are kept same as before. As usual validation set was in parameter tuning.  The original Audioset gives label density at 10 seconds segment. Hence, we are analyzing the performance of models trained on datasets for which weak labels are available at 10 seconds, 30 seconds and 60 seconds granularity.  

Table \ref{tab:labelden} shows the effect of label density change in MAP and MAUC numbers. The MAP value decreases by a sharp $\mathbf{12\%}$ in relative terms as we move from weak labels at 10 seconds long audio recordings in Audioset to 30 seconds long audio recordings in Audioset-At-30. The drop in performance is evident in both MAP and MAUC numbers. However, going from 30 to seconds leads to an additional drop of only $\mathbf{4\%}$ for MAP. No significant change in MAUC is observed. This shows an interesting behavior. As we increase the recording durations from 10 to 30 seconds, the model is not able to learn well from relatively reduced label density. However, increasing audio lengths also increases the total amount of training data and one might attribute this behavior to the increased training data. Larger training data have been known to improve the robustness of deep learning models. This characteristic of deep learning methods starts showing up when we train on Audioset-At-60. Although, the performance does go down due to reduced label density, training on larger amount of data does improve the robustness of WAL-Net and the drop in performance is restricted to a certain extent. Hence, when weak label learning is done on long duration recordings, it is desirable to use large amounts of data to counter the effects of increased label density noise. 

\begin{table}[t!]
  \centering
  \caption{Effect of label density on performance   }
    \begin{tabular}{|l|l|l|}
    \hline
    \textbf{Training Set} & \textbf{MAP} & \textbf{MAUC} \\
    \hline
    Audioset  & 0.196 & 0.925 \\
    \hline
    Audioset-At-30 & 0.172 (\textbf{-12\%}) & 0.904 (-2\%)\\
    \hline
    Audioset-At-60 & 0.165 (\textbf{-16\%}) & 0.908 (-2\%) \\
    \hline
    \end{tabular}%
  \label{tab:labelden}%
\end{table}%
\begin{table}[t!]
  \centering
  \caption{10 Audio Events with highest drop in AP for Audioset-At-30 (w.r.t Audioset)  }
\resizebox{1.0\columnwidth}{!}{     \begin{tabular}{|l|l|l|l|l|}
    \hline
    \textbf{Events}  & \textbf{Audioset - AP, (AUC)} & \textbf{Audioset-At-30 - AP, (AUC)} & \textbf{AP Drop (\% Drop)} \\
    \hline
    Speech synthesizer & 0.225, (0.953) & 0.037, (0.882) & 0.187, (83.506) \\
    \hline
    Air horn truck horn & 0.290, (0.979) & 0.051, (0.950) & 0.239, (82.310) \\
    \hline
    Vehicle horn, honking & 0.238, (0.955) & 0.084, (0.923) & 0.155, (64.913) \\
    \hline
    Afrobeat & 0.261, (0.977) & 0.104, (0.967) & 0.157, (60.088) \\
    \hline
    Whip & 0.274, (0.927) & 0.112, (0.876) & 0.162, (59.117) \\
    \hline
    Chopping (food) & 0.285, (0.920) & 0.117, (0.886) & 0.168, (58.887) \\
    \hline
    Music of Bollywood & 0.309, (0.933) & 0.140, (0.933) & 0.169, (54.765) \\
    \hline
    Sizzle & 0.339, (0.976) & 0.155, (0.958) & 0.183, (54.122) \\
    \hline
    Toot & 0.290, (0.963) & 0.137, (0.961) & 0.153, (52.813) \\
    \hline
    Pour & 0.238, (0.957) & 0.115, (0.940) & 0.123, (51.717) \\
    \hline
    \textbf{Mean} & \textbf{0.275, (0.954)} & \textbf{0.105, (0.927)} & \textbf{0.170, (62.224)} \\
    \hline
    \end{tabular}%
    }
  \label{tab:labelden30}%
  \vspace{-0.1in}
\end{table}%

\begin{table}[t!]
  \centering
  \caption{10 Audio Events with highest drop in AP for Audioset-At-60 (w.r.t Audioset)  }
  \resizebox{1.0\columnwidth}{!}{    \begin{tabular}{|l|l|l|l|l||}
    \hline
    \textbf{Sound Events}  & \textbf{Audioset - AP, (AUC)} & \textbf{Audioset-At-60 - AP, (AUC)} & \textbf{AP Drop (\% Drop)} \\
    \hline
    Chopping (food) & 0.285, (0.920) & 0.064, (0.871) & 0.220, (77.348) \\
    \hline
    Pink noise & 0.349, (0.974) & 0.093, (0.942) & 0.257, (73.448) \\
    \hline
    Cash register & 0.205, (0.917) & 0.056, (0.918) & 0.149, (72.722) \\
    \hline
    Whip & 0.274, (0.927) & 0.078, (0.877) & 0.196, (71.705) \\
    \hline
    Music of Africa & 0.225, (0.966) & 0.082, (0.919) & 0.143, (63.512) \\
    \hline
    Keyboard (musical) & 0.303, (0.955) & 0.116, (0.907) & 0.186, (61.524) \\
    \hline
    Afrobeat & 0.261, (0.977) & 0.102, (0.955) & 0.159, (61.038) \\
    \hline
    Speech synthesizer & 0.225, (0.953) & 0.089, (0.905) & 0.136, (60.401) \\
    \hline
    Electronic organ & 0.205, (0.916) & 0.089, (0.847) & 0.116, (56.491) \\
    \hline
    Artillery fire & 0.218, (0.963) & 0.102, (0.939) & 0.117, (53.482) \\
    \hline
    \textbf{Mean} & \textbf{0.255, (0.947)} & \textbf{0.087, (0.908)} & \textbf{0.168, (65.167)} \\
    \hline
    \end{tabular}%
    }
  \label{tab:labelden60}%
  \vspace{-0.1in}
\end{table}%

Table \ref{tab:labelden30} shows 10 events for which maximum drop (relative) in AP is observed for Audioset-At-30 compared to Audioset. For the purpose of showing events in Table \ref{tab:labelden30}, we considered only those whose AP is more than the MAP (0.196). Similarly, Table \ref{tab:labelden60} shows the same for Audioset-At-60. For these 10 events, on an average around 62\% drop in AP is seen for Audioset-At-30 and 65\% for Audioset-At-60. Note that the set of events for which maximal drop is observed for Audioset-At-30 and Audioset-At-60 are different. Interestingly, most of the events in these two lists, except for \emph{Pink Noise, Music of Bollywood, Music Africa} have unique and specific characteristics and are easily identifiable by humans. One can then argue that such events are more prone to suffer from label density noise.  

\subsection{Analysis of Labels Corruption}
\begin{figure}[t]
  \includegraphics[width=0.52\textwidth]{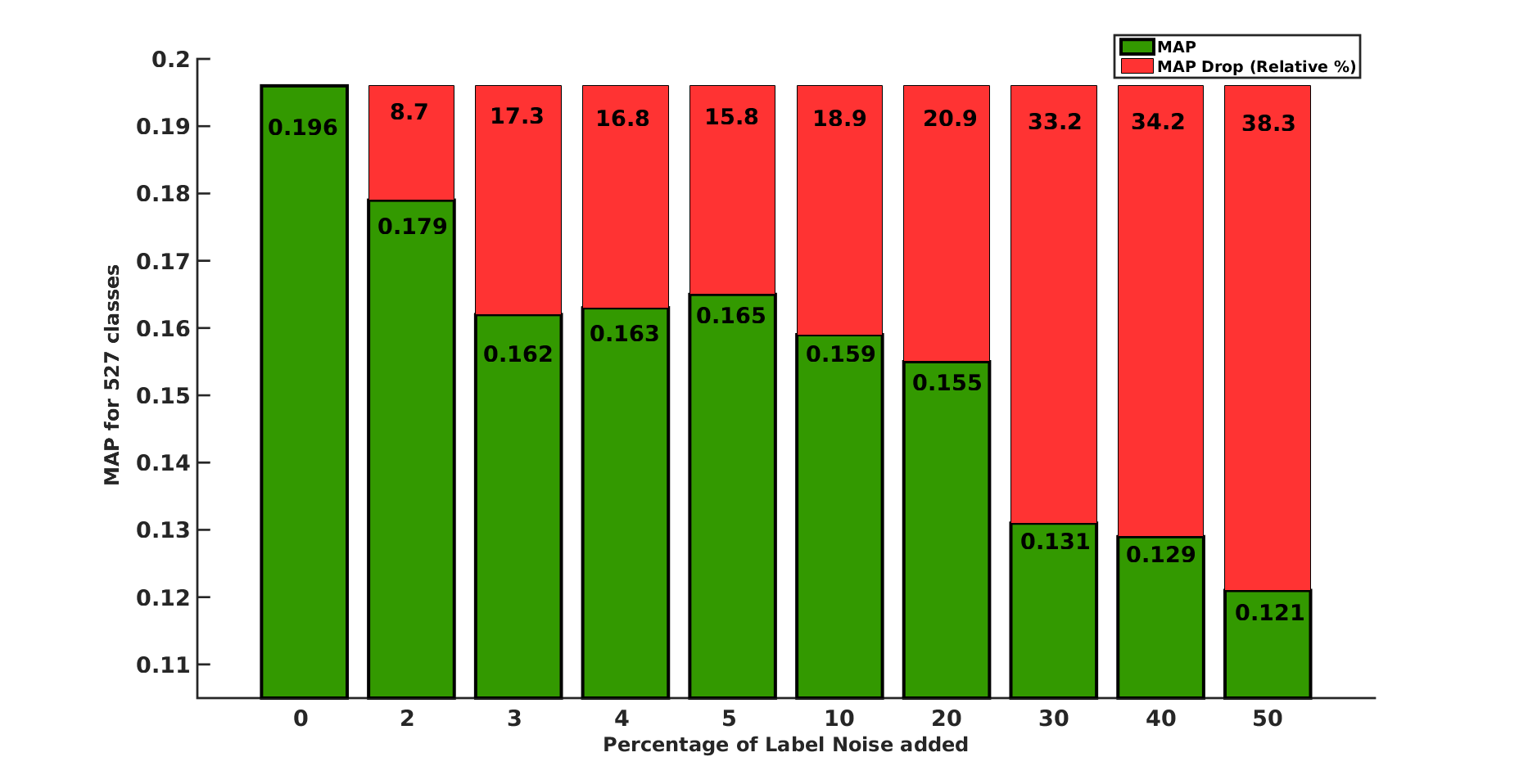}
  \caption{Effect of corruption of labels on performance. x-axis represents corruption level $r$ in \%. y-axis show MAP and \% reduction in MAP compared to no corruption.}
  \label{fig:labelcor}
  \vspace{-0.1in}
\end{figure}
The analysis of label noise in terms of corrupted labels follows the procedure outlined in \ref{ssec:labcor}. We use 9 different values of $r$, the level of artificially induced corruptions in labels. Correspondingly, 9 different training sets are obtained. A WAL-Net is trained on each of these training sets. The validation and evaluation sets remain as before. Fig. \ref{fig:labelcor} shows MAP values for different values of $r$. $r=0$ corresponds to the original training set from Audioset. 

Adding around $2\%$ noise in labels leads to around $8.7\%$ drop in MAP performance. The model remains fairly robust to additional noise till $r=5$. In fact increasing noise labels to around $20\%$ noise. $r=20\%$ leads to an extra drop of only $4\%$ in MAP, compared to the performance at $r=4\%$. This shows that WAl-Net is robust to a certain extent to label noise. At very high $r=40$ and $r=50$, as expected a sharp decrease in performance is noted. The MAP at these $r$ are $12.93 (-34\%)$ and $12.1(-38\%)$. 

Overall, we can state that corrupted labels are important and depending on the amount of label noise, a considerable drop in performance can be observed. Still, a deep learning model like WAL-Net seems robust to a certain extent. Future works on AED using weak labels, especially those relying on YouTube data should factor in the possibility of corrupted labels and model accordingly. As pointed out earlier, even manually labeled dataset like Audioset is not free of label noise and it is expected that any other large scale weakly labeled dataset will have the same problem. Hence, training algorithms should be built around label noise in datasets. 

Table \ref{tab:labelnoise30} shows top 10 events for which maximum relative drop in performance is observed. The drop of around $90\%$ for \emph{Firecracker} is especially noticeable. On an average these 10 events show a drop of $77\%$ in performance. There is drop of around 10.5\% in AUC values as well. Note that two events, \emph{Cattle bovine} and \emph{Moo} are expected to often co-occur together. For these two events we have almost similar performance and the performance after noise addition is also similar. This is consistent with the design of our labels corruption strategy. 
\begin{table}[t!]
  \centering
  \caption{10 Events with highest drop in AP (for $r=30\%$ label noise)  }
\resizebox{1\columnwidth}{!}{   \begin{tabular}{|l|l|l|l|}
    \hline
    \textbf{Events} & \textbf{AP, AUC @ No Noise} & \textbf{AP, AUC @ 30 \% Noise} & \textbf{Drop in AP, (\% Drop )} \\
    \hline
    Firecracker & 0.223, (0.964) & 0.023, (0.814) & 0.200, (89.737) \\
    \hline
    Cough & 0.218, (0.952) & 0.038, (0.831) & 0.180, (82.771) \\
    \hline
    Chainsaw & 0.395, (0.947) & 0.079, (0.861) & 0.316, (79.995) \\
    \hline
    Acoustic guitar & 0.259, (0.967) & 0.052, (0.857) & 0.207, (79.972) \\
    \hline
    Waterfall & 0.224, (0.975) & 0.054, (0.918) & 0.170, (75.737) \\
    \hline
    Car  & 0.214, (0.921) & 0.056, (0.751) & 0.158, (73.755) \\
    \hline
    Sizzle & 0.339, (0.976) & 0.091, (0.896) & 0.248, (73.049) \\
    \hline
    Cattle bovine & 0.214, (0.930) & 0.059, (0.854) & 0.155, (72.506) \\
    \hline
    Moo  & 0.219, (0.923) & 0.061, (0.880) & 0.159, (72.400) \\
    \hline
    Lawn mower & 0.204, (0.972) & 0.058, (0.856) & 0.146, (71.685) \\
    \hline
    \textbf{Mean} & \textbf{0.251, (0.953)} & \textbf{0.057, (0.852)} & \textbf{0.194, (77.161)} \\
    \hline
    \end{tabular}%
    }
  \label{tab:labelnoise30}%
\end{table}%

\subsection{Weakly Labeled Audio in the Wild}
\begin{table}[t!]
  \centering
  \caption{Weakly Labeled Audio in the Wild. Comparison with Audioset}
    \begin{tabular}{|l|l|l|}
    \hline
    \textbf{Training Set} & \textbf{MAP} & \textbf{MAUC} \\
    \hline
    Audioset-40  & 0.419 & 0.913 \\
    \hline
    YouTube-Wild & 0.127 & 0.70- \\
    \hline
    \end{tabular}%
  \label{tab:ytwild}%
   \vspace{-0.1in}
\end{table}%

In this Section, we analyze the situation in which the previous two forms of ``label noise" occurs naturally and in fact in abundance, which is going to affect the learning process. We work directly with audio (video) obtained from YouTube and labeled without any manual intervention. The procedure follows the process outlined in Section \ref{ssec:ytwild}. As noted in Section \ref{ssec:ytwild}, we work with only a small subset of events from Audioset, events with salient characteristics and ``names" which can be used in query terms to yield at least somewhat relevant results from YouTube.

\begin{table}[t!]
  \centering
  \caption{Best 10 and Worst 10 performing events (ordered by AP) for YouTube-Wild}
   \resizebox{1\columnwidth}{!}{     \begin{tabular}{|l|l|l|l|l|l|}
    \hline
    \textbf{Events (Best 10)} & \textbf{YouTube-Wild} & \textbf{Audioset-40} & \textbf{Events (worst 10)} & \textbf{YouTube-Wild} & \textbf{Audioset-40}\\
    \hline
    Guitar & 0.316, (0.748) & 0.755, (0.961) & Engine & 0.048, (0.488) & 0.330, (0.896) \\
    \hline
    Siren & 0.291, (0.731) & 0.735, (0.970) & Rail transport & 0.042, (0.564) & 0.441, (0.929) \\
    \hline
    Animal & 0.289, (0.712) & 0.579, (0.882) & Violin, fiddle & 0.041, (0.692) & 0.446, (0.945) \\
    \hline
    Chicken, rooster & 0.268, (0.824) & 0.281, (0.908) & Tools & 0.039, (0.540) & 0.356, (0.872) \\
    \hline
    Vehicle & 0.243, (0.566) & 0.482, (0.828) & Bus  & 0.039, (0.779) & 0.052, (0.818) \\
    \hline
    Emergency vehicle & 0.234, (0.728) & 0.613, (0.961) & Train & 0.035, (0.476) & 0.427, (0.925) \\
    \hline
    Laughter & 0.233, (0.881) & 0.612, (0.960) & Motorboat, speedboat & 0.035, (0.732) & 0.060, (0.845) \\
    \hline
    Drum & 0.224, (0.747) & 0.572, (0.948) & Truck & 0.032, (0.512) & 0.134, (0.878) \\
    \hline
    Drum kit & 0.218, (0.855) & 0.530, (0.974) & Race car, auto racing & 0.030, (0.622) & 0.157, (0.884) \\
    \hline
    Crowd & 0.191, (0.693) & 0.681, (0.979) & Motorcycle & 0.026, (0.599) & 0.066, (0.826) \\
    \hline
    \textbf{Mean} & \textbf{0.251, (0.748)} & \textbf{0.584, (0.937)} & \textbf{Mean} & \textbf{0.037, (0.600)} & \textbf{0.247, (0.882)} \\
    \hline
    \end{tabular}%
    }
  \label{tab:ytwildtp}%
   \vspace{-0.1in}
\end{table}%
We selected a total of $40$ sound events and obtained training examples for them from YouTube by the process described in Section \ref{ssec:ytwild}. This set, called \emph{YouTube-Wild} is used as the training set for training WAL-Net. It consists of a total of 1906 videos for $40$ sound events, totaling around 60 hours of audio. The average duration of audio recordings is around $111$ seconds, with  maximum duration of $240$ seconds. In comparison, Audioset has mostly 10 seconds long audio recordings. All audio recordings are sampled to $44.1$ kHz sampling rate. \emph{YouTube-Wild} training list will be released and made available for future works in this area.  

We train WAL-Net (with $C_s = 40$) on YouTube-wild. We also train WAL-Net ($C_s = 40$) on the subset Audioset training set, which contains audio recordings belonging to only these 40 sound events. We call this subset of of Audioset training set as \emph{Audioset-40}. The subset of Audioset validation and evaluation sets which contains only recordings belonging to only these 40 events are used as validation and evaluation set in these experiments.   

Table \ref{tab:ytwild} compares performance on Audioset-40 and YouTube-Wild. One can observe the considerable difference between the performance of these training sets. Audioset-40 is manual labeled dataset and is also expected to have reasonably good label density as recordings are of 10 seconds duration. On the other hand, YouTube-wild is not manually labeled or even verified and is expected to have large number corrupted labels. Moreover, YouTube-wild consists of very long duration recordings (upto 4 minutes) and hence even those where the weak labels are correct, label density noise can be very high. Together these two forms of noises severely affects the learning process. Even though WAL-Net showed reasonably good robustness in the previous sections, we observe that learning sound events from wild web data without any manual intervention is an extremely difficult task. Future works on large scale AED needs to develop algorithms which can address these challenges. 

Table \ref{tab:ytwildtp} shows 10 best and worst performing events for YouTube-Wild. Corresponding Audioset-40 numbers for these events are shown the Table.  Even for the 10 event where YouTube-Wild does really well, compared to Audioset-40 an average drop of 55\% in AP is noted. At the same time, there are a few events such as \emph{Chicken, Rooster} among best 10 and \emph{Bus}, \emph{Motorboat} and \emph{Motorcycle} among worst 10 where both sets perform equally well or not. 

Table \ref{tab:ytwilddrop} shows 10 events for which highest percentage drop in performance is observed. Corresponding AUC values are also shows in parenthesis. As expected the average drop over these cases is very high, greater than 80\%. Interestingly, 3 of the classes (\emph{Train, Rail Transport, and Railroad Car-Wagon}) are broadly expected to fetch similar results from YouTube. For all of these 3 classes performance drop is large. 
\begin{table}[t!]
  \centering
  \caption{10 Events with highest relative drop in performance  }
 \resizebox{0.98\columnwidth}{!}{
 	\begin{tabular}{|l|l|l|l|}
    \hline
    \textbf{Events} & \textbf{Audioset-40 - AP (AUC)} & \textbf{YouTube-Wild} & \textbf{AP Drop (\% Drop)} \\
    \hline
    Train & 0.427, (0.925) & 0.035, (0.476) & 0.392, (91.861) \\
    \hline
    Violin fiddle & 0.446, (0.945) & 0.041, (0.692) & 0.405, (90.727) \\
    \hline
    Rail transport & 0.441, (0.929) & 0.042, (0.564) & 0.399, (90.386) \\
    \hline
    Singing & 0.659, (0.947) & 0.089, (0.529) & 0.569, (86.420) \\
    \hline
    Keyboard (musical) & 0.568, (0.949) & 0.084, (0.735) & 0.484, (85.168) \\
    \hline
    Railroad car,  train wagon & 0.451, (0.934) & 0.071, (0.697) & 0.380, (84.313) \\
    \hline
    Water & 0.616, (0.935) & 0.118, (0.621) & 0.498, (80.817) \\
    \hline
    Bass drum & 0.541, (0.973) & 0.119, (0.782) & 0.422, (78.070) \\
    \hline
    Cymbal & 0.577, (0.977) & 0.151, (0.833) & 0.426, (73.911) \\
    \hline
    Pigeon, dove & 0.526, (0.956) & 0.140, (0.771) & 0.386, (73.388) \\
    \hline
    \textbf{Mean} & \textbf{0.525, (0.947)} & \textbf{0.089, (0.670)} & \textbf{0.436, (83.506)} \\
    \hline
    \end{tabular}%
    }
  \label{tab:ytwilddrop}%
   \vspace{-0.1in}
\end{table}%
\section{Conclusion}
\label{sec:conclusions}
In this paper, we attempted to understand the challenges in large scale AED using weakly labeled data. We first described a CNN based framework to learn from weak labeled audio recordings. The network design embodies several desirable characteristics from the perspective of weakly labeled audio event detection. It can handle recordings of variable length, the network design controls segment sizes over which secondary outputs are produced and can be adjusted. This removes any additional pre-processing step. The network is able to achieve state-of-art performance (using balanced training set) of Audioset. 

We then described different ways in which label noise manifests itself in weakly labeled data. We proposed experimental designs which can help study the effect of these factors. In particular, we looked into how \emph{label density} and \emph{labels corruption} affects performance. Moreover, we also considered directly mining weakly labeled data from web and compared it with Audioset, which has been manually labeled over shorter audio recordings. The primary motivation behind these is to understand the challenges and help future works on AED with weak labels adapt accordingly. Mining weakly labeled data from web is not easy and one needs to bring in sophisticated algorithms to assign weak labels to the audios. Label density as described in this work naturally comes in weak label learning paradigm. Algorithms should be designed to handle audio recordings with wide range of label densities. Corruption in noise labels is another important factor and again robustness to a certain extent again labels corruption is desirable.

\bibliographystyle{IEEEtran}
\bibliography{final}


%








\end{document}